\documentclass[12pt,twoside]{article}

\date{}
\usepackage {graphicx} 
\def\2F1{~_2F_1}

\begin{document}
\centerline{\bf Adv. Studies Theor. Phys, Vol. x, 200x, no. xx,
xxx - xxx}

\centerline{}

\centerline{}
\centerline {\Large{\bf 
A  law of motion 
for 
spherical shells  
 }}

\centerline{}

\centerline{\Large{\bf in special relativity 
}}

\centerline{}

\centerline{\bf {L. Zaninetti}}

\centerline{}

\centerline{Dipartimento  di Fisica Generale,}

\centerline{Universit\`a degli Studi di Torino,}

\centerline{via P. Giuria 1,  10125 Torino,Italy}

\begin{abstract}
Self-similar solutions to the problem of 
a  special relativistic law of motion 
for  thin shells of matter are calculated. 
These solutions represent the special 
relativistic generalization 
of momentum conservation for the thin layer 
approximation in classical physics.
The analytical and numerical results are 
applied to Supernova Remnant 1987~A.

\end{abstract}


\section  {Introduction}
The study of the relativistic dynamics of thin shells 
of matter is a current subject of investigation
when the framework of general relativity
is adopted \cite{Israel1966,Israel1967,Nunez1996,Kijowski2006}.
Here, we will explore how the framework of special
relativity can produce a law of motion
which can be compared with expansion data
of a supernova remnant (SNR).
From a classical point of view, 
the temporal evolution  of the SNR 
is modelled by adopting different 
physical approaches that are not invariant under relativistic
transformations.
As an example,  
a  classical  analytical solution
of the Sedov 
type \cite{Sedov1946,Taylor1950,Sedov1959} 
is equation~(10.27) in 
\cite{Dalgarno1987}
\begin{equation}
R(t)= \bigl ( \frac {25 E t^2 }
              {4 \pi \rho } \bigr )^{1/5}
\label{eq:radiussnr}
\quad,
\end{equation}
where $\rho$  is the density of the surrounding  medium
which is supposed to be constant, $E$ is the energy of
the explosion
and  $t$ is the age  of the SNR.
In this  paper, we show that it is 
possible to  deduce  the relativistic  law of motion
starting from the  conservation of relativistic momentum.
The relativistic equation of a SNR is deduced as 
a non-linear relationship between radius and time 
which can be solved numerically.
The self-similar relativistic solutions are accurate
for sufficiently large time after the formation of the SNR.
The theory is applied to the SNR connected
with Supernova (SN) 1987A.

\section {Classical and relativistic laws of motion}

The thin layer approximation 
in classical physics 
assumes that all the swept-up 
gas accumulates with infinite density in a thin shell just after
the shock front.
The conservation of radial momentum requires that,
after the initial  radius $R_0$,  
\begin{equation}
\frac{4}{3} \pi R^3 \rho V = 
\frac{4}{3} \pi R_0^3 \rho V_0
\quad ,
\end{equation}
where $R$ and $V$   are  the radius and velocity
of the advancing shock wave,
$\rho$ is the density of the ambient medium
and 
$V_0$ is the  initial velocity,
see \cite{Dyson1997,Padmanabhan_II_2001}.
In classical physics, 
the velocity as a function of radius
is: 
\begin{equation}
\label{velocityclassical}
V=  V_0 (\frac {R_0}{R})^3 
\quad , 
\end{equation}
the law of motion is: 
\begin{equation}
\label{radiusclassical}
R = R_0 \left  ( 1 +4 \frac{V_0 } {R_0}(t-t_0) \right )^{\frac{1}{4}}  
\quad ,
\end{equation}  
where $t$ is time and $t_0$ is
the initial  time.
In classical physics, 
the velocity as a function of time
is: 
\begin{equation}
V  =  {V_0} \left ( 1 +4 \frac{V_0} {R_0}(t-t_0)\right )^{-\frac{3}{4}}  
\label{velocitym} 
\quad . 
\end{equation}   
Equation (\ref{velocityclassical}) can  also be solved 
 with a similar solution of type $R=K(t-t_0)^{\alpha}$,
$k$ being a constant,  
and the classical result is:  
\begin{equation}
\label{radiussimilar}
R =  \sqrt [4]{4}\sqrt [4]{\beta_{{0}}{R_{{0}}}^{3}c}
 (t-t_0)^{\frac{1}{4}}  
\quad ,
\end{equation}  
where  $\beta_0= \frac {V_0}{c} $ has  been introduced
in order to make a comparison with the relativistic
case. 

Newton's law in special relativity is: 
\begin{equation}
F = \frac {dp} {dt } = \frac {d} {dt} ( m V) 
\quad ,
\end{equation}
where $F$ is the force,
$p$ the relativistic momentum,
$m$ the relativistic mass,
$m_0$ the mass at rest and 
$V$ the velocity,
see  equation~(7.16) in~\cite{French1968}.  
In the case of the relativistic expansion of a shell 
in which all the swept material 
resides at two different points,  
denoted by radius $ R $  and $R_0$,
 we have:
\begin{equation}
\frac 
{\rho \frac {4}{3} \pi R^3 \beta } 
{\sqrt  {1-\beta^2}}
=
\frac 
{\rho \frac {4}{3} \pi R_0^3 \beta_0 } 
{\sqrt  {1-\beta_0^2}}
\quad ,
\end{equation}
where  
$\beta_0$=$V_0/c$,
$\beta$=$V/c$ 
and
$c$ is  the velocity of light.
The velocity  of the relativistic expanding shell is: 
\begin{equation}
\label {velocityrelativistic}
\beta=
\frac
{
{R_{{0}}}^{3}\beta_{{0}}
}
{
\sqrt {{R_{{0}}}^{6}{\beta_{{0}}}^{2}+{R}^{6}-{R}^{6}{\beta_{{0}}}^{2}
}
}
\quad .
\end{equation}

\begin{figure}
  \begin{center}
\includegraphics[width=8cm]{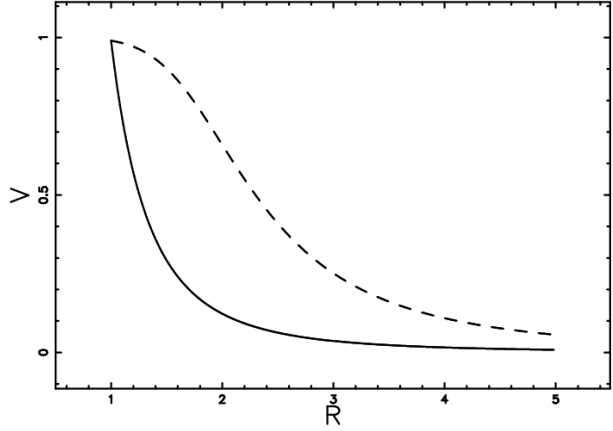}
  \end {center}
\caption {
 Velocity as a function of radius 
 when  $R_0=1$, $c=1$,
 $v_0/c =0.99$.
 ~(relativistic~case,~equation~(\ref{velocityrelativistic})) 
       (dashed-line        )
 and 
 ~(classical ~case,~equation~(\ref{velocityclassical})) 
       (full-line        )
          }%
    \label{velocity}
    \end{figure}

Figure~\ref{velocity} shows the classical and relativistic
behaviors of the velocity as a function of radius $R$.

The previous formula can be expressed in differential 
form as:
\begin{equation}
\sqrt {{R_{{0}}}^{6}{\beta_{{0}}}^{2}+{R}^{6}-{R}^{6}{\beta_{{0}}}^{2}
}{\it dR}
=
{R_{{0}}}^{3}\beta_{{0}}c{\it dt}
\quad  .
\label{integrale}
\end{equation}
The integral on the lhs of the previous equation
can be evaluated with a first transformation
$\mu ={\frac {1}{{R_{{0}}}^{6}{\beta_{{0}}}^{2}}}-\frac{1}{R_0^6} $ and
$x=R$ ,
\begin{equation}
\int \!\sqrt {1+\mu\,{x}^{6}}{dx}
\quad .
\end{equation}
A second transformation $y=x^6$ changes the integral 
into: 
\begin{equation}
\int \frac{1}{6} \,{\frac {\sqrt {1+\mu\,y}}{{y}^{5/6}}}{dy}
\quad .
\end{equation}
This  integral is of the same type 
as formula 3.194.1 in 
\cite{Gradshteyn2007}
\begin{equation}
\int _{0}^{u}\!{\frac {{x}^{\mu-1}}{ \left( 1+\beta\,x \right) ^{\nu}}
}{dx}=
{\frac {{u}^{\mu}{\2F1(\mu,\nu;\,1+\mu;\,-\beta\,u)}}{\mu}}
\quad ,
\end{equation}
where ${\2F1(a,b;\,c;\,z)}$ 
is a regularized hypergeometric function
\cite{Abramowitz1965,Seggern1992,Thompson1997,Gradshteyn2007}.
In our case, 
$\nu =\frac{1}{2}$ and  $\mu=\frac{1}{6}$.
The hypergeometric function 
is defined 
by the following power series expansion:  
\begin{equation}
_2F_1(a,b;\,c\,;z)=\sum_{n=0}^{\infty} \frac{(a)_n(b)_n}{(c)_{n}n!}z^n
\quad  ,
\end{equation}
where 
$(w)_n$
is the Pochhammer symbol
\begin{equation}
(w)_n=w(w+1)\dots(w+n-1)
\quad  ,
\end{equation}
 $(a,b,c)$  is a triplet of real numbers 
with $c$ not belonging to the
set of negative integers and $z$ is a real number $ < 1$. 

We are now ready 
to integrate formula~(\ref{integrale} )
and the result is 
a  non-linear equation, 
 ${\mathcal{F}}_{NL}$, in $R$: 
\begin{eqnarray}
\label{radiusrelativistic}
{\mathcal{F}}_{NL} =  \nonumber \\ 
 R
{\2F1(-1/2,1/6;\,7/6;\,{\frac {{R}^{6} \left( -1+{\beta_{{0}}}^{2} \right)
}{{R_{{0}}}^{6}{\beta_{{0}}}^{2}}})}
\nonumber \\
-R_{{0}}
{\2F1(-1/2,1/6;\,7/6;\,{\frac {-1+{\beta_{{0}}}^{2}}{{\beta_{{0}}}^{2}}})}
\nonumber \\
-c ( t-t_{{0}} )   {R_{{0}}}^{3}\beta_{{0}}=0
\end{eqnarray}

\begin{figure}
\begin{center}
\includegraphics[width=8cm]{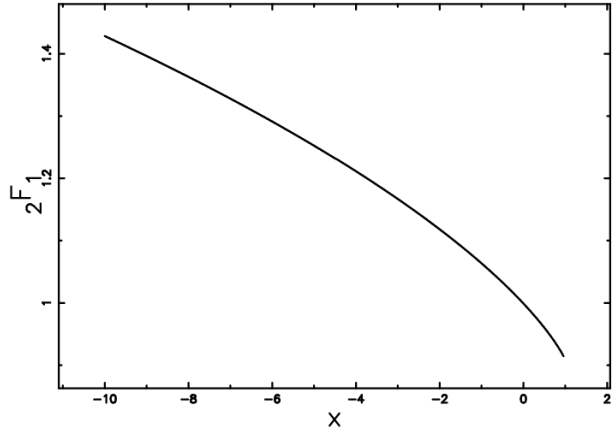}
\end  {center}
\caption {
Plot of the hypergeometric function
  $\2F1(-1/2,1/6;\,7/6;\,x)$ as a function of $x$
 (full line) 
for  $-10 \leq x <  +1 $ ~.
}
\label{hyper}
\end{figure}

From a numerical  point of view,  
the hypergeometric function can be evaluated
 with 
the FORTRAN subroutine  HYGFX 
extracted  from~\cite{Zhang1996} and a
typical  plot  of 
$\2F1(1/6,1/2;\,7/6;\,x)$ is reported
in Figure~\ref{hyper}.

\begin{figure}
  \begin{center}
\includegraphics[width=8cm]{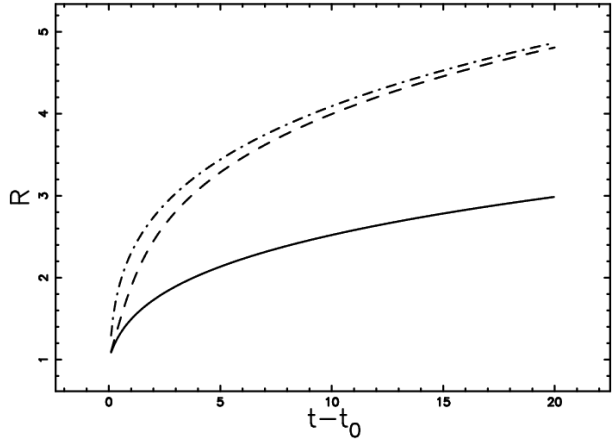}
  \end {center}
\caption {
 Radius  as a function of time  
 when  $R_0=1$, $\beta_0=0.99 $,
 $c=1$, 
   (asymptotic relativistic~case,~equation~(\ref{radiusrelativisticasyn})) 
   (dot-dash-dot-dash line),
   (relativistic ~case,~equation~(\ref{radiusrelativistic})) 
   (dashed-line           )
and  
(classical ~case,~equation~(\ref{radiusclassical}))(full-line) 
          }%
    \label{radius}
    \end{figure}

\begin{figure}
  \begin{center}
\includegraphics[width=8cm]{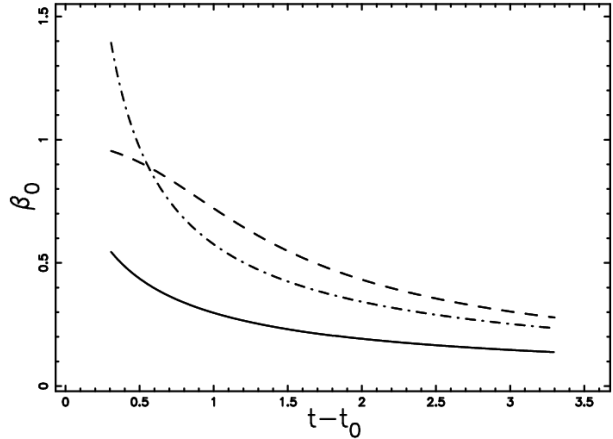}
  \end {center}
  \caption {
 Velocity   as a function of time  
 when  $R_0=1$, $\beta_0=0.99 $ and  
 $c=1$.  
 The dashed-line represents 
 the  relativistic case,
 the dot-dash-dot-dash represents the 
 asymptotic relativistic case 
 and  
 the full line the classical case. 
          }%
    \label{vel_time}
    \end{figure}

Once the  hypergeometric function   is implemented,
we can solve the non-linear equation~(\ref {radiusrelativistic})
with the FORTRAN subroutine  ZRIDDR 
from \cite{press} and a typical example
is shown in Figure~(\ref{radius}).
Once  a numerical expression for the relativistic radius
is obtained, we can easily  obtain
a velocity-time relationship from 
equation~(\ref{velocityrelativistic}), see 
Figure~\ref{vel_time}.

An approximate solution of 
equation~\ref{velocityrelativistic} can be found by imposing
$R(t)=k(t-t_0)^{\alpha}$. 
The self-similar solutions  of the relativistic case
under the assumption  $R^6 (1-\beta_0^2) \gg 
R_0^6\beta_0^2$  
are:
\begin{equation}
R(t)=
\sqrt {2}\sqrt [4]{{\frac {\beta_{{0}}{R_{{0}}}^{3}c}{\sqrt {1-{\beta_
{{0}}}^{2}}}}}\sqrt [4]{(t-t_0)}
\quad ,  
\label{radiusrelativisticasyn}
\end{equation}
and
\begin{equation}
\beta(t)=
1/4\,\sqrt {2}\sqrt [4]{{\frac {\beta_{{0}}{R_{{0}}}^{3}c}{\sqrt {1-{
\beta_{{0}}}^{2}}}}}{(t-t_0)}^{-3/4}{c}^{-1}
\quad .
\label{velocityrelativisticasyn}
\end{equation}

\section {Application to SN 1987A}

\begin{figure}
  \begin{center}
\includegraphics[width=8cm]{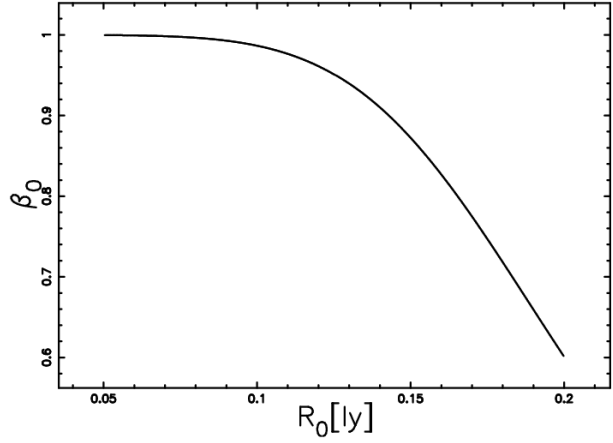}
  \end {center}
\caption 
{
Initial  velocity $\beta_0$ as a function of 
initial radius when $R=0.61~ly$, $t=21.86~yr$ 
and $c=1$.
}
    \label{initials}
    \end{figure}

\begin{figure}
  \begin{center}
\includegraphics[width=8cm]{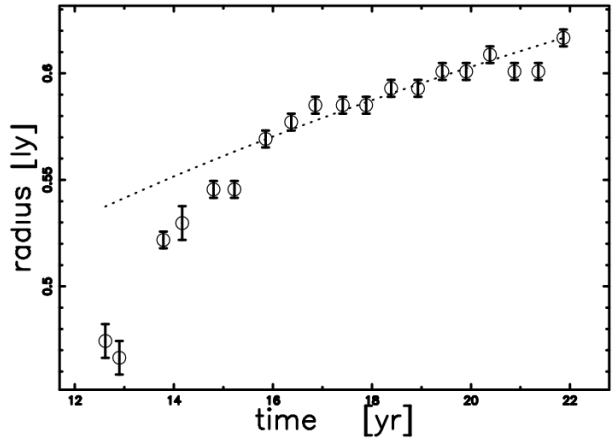}
  \end {center}
\caption 
{
 Radius  as a function of time  
 when  $R_0=0.079$, $\beta_0=0.95 $,
 $c=1$ 
   (asymptotic relativistic~case,~equation~(\ref{radiusrelativisticasyn})) 
   (dotted line)
   with the addition of the  observed 
   radius with relative error extracted 
   from  \cite{Park2007}
          }%
    \label{realevolution}
    \end{figure}

The SN 1987A exploded in the Large Magellanic Cloud 
in 1987.
The distance of this SN is $\approx~50~kpc$ 
(163050~$ly$) and a detailed analysis
of the distance, $D$,
gives 
$D=51.4~kpc$  \cite{Panagia2005}
and
$D=50.18~kpc$  \cite{Mitchell2002}.
After 7987~$days$, the diameter of the SNR
was  $0.77^{\prime\prime}$ 
and it's velocity $\approx~1412~km/s$ 
\cite{Racusin2006}\cite{Park2007}.
In this Section we will adopt year ($yr$) as a time
unit and light year ($ly$) as a length unit;
with these natural units $c=1$.
The radius of  the SNR   after 21.86 $yr$ is
\begin{equation}
R =r_{77} \times D_{50} 0.61~ly 
\quad ,
\end{equation}
where $r_{77}$ is the radius in arcsec divided by 0.77 
and   $D_{50}$ the distance in pc divided by 50000.
Next, we attempt to evaluate the initial conditions 
$R_0$ and $\beta_0$ after $t=21.86~yr$ and
$R=0.61~ly$ are given.
The approximate self-similar relativistic solution 
for the radius as represented by 
equation (\ref{radiusrelativisticasyn}) allows 
a relationship to be determined 
between  these  two unknown variables,
\begin{equation}
\beta_0 =
{\frac {{R}^{4}}{\sqrt {{R}^{8}+16\,{R_{{0}}}^{6}{c}^{2}{t}^{2}}}}
\quad ,
\label{iniziali}
\end{equation}
and  Figure~\ref{initials} reports such correlated 
initial conditions.
Figure~\ref{realevolution} shows the behavior of the 
radius as a function 
of time  
after setting initial conditions given by 
equation~(\ref{iniziali}) 
and extracting
the data of  SNR~1987A 
from Figure~2 in \cite{Park2007}.

\begin{figure}
  \begin{center}
\includegraphics[width=8cm]{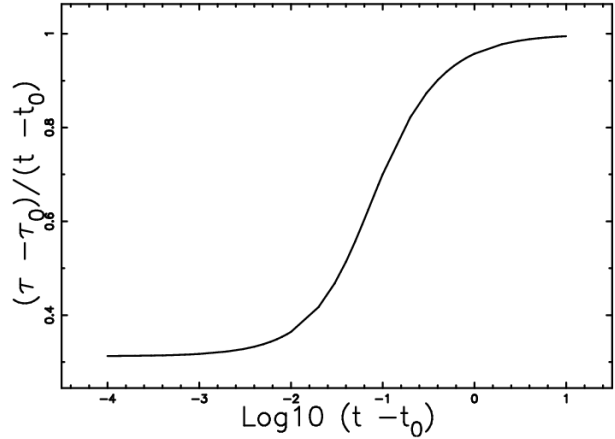}
  \end {center}
\caption 
{
 The time contraction  $r$ of the clock on the 
 expanding surface 
 as a function of the decimal logarithm
 in laboratory time when  $R_0=0.079$, $\beta_0=0.95 $
 and $c=1$.
          }%
    \label{contraction}
    \end{figure}

A further test can be done by inserting  in 
formula~(\ref{velocityrelativisticasyn}) for the 
self-similar relativistic velocity 
the boundary conditions used to deduce the trajectory,
i.e. $R_0=0.079$, $\beta_0=0.95$ and $t=21.86~yr$;
the theoretical velocity turns out to  be 
$2672~\frac{km}{s}$ against the observed 
$1412~\frac{km}{s}$ \cite{Park2007}.
The proper time of the world line 
is
\begin{equation}
\tau - \tau_0 =
\int_{\tau_0}^{\tau} 
\sqrt { 1 -\beta(t)^2}  dt
\quad .
\end {equation}
This means a relativistic time contraction, $r$,
for the clock which follows the expansion 
\begin{equation}
r  = \frac {\tau -\tau_0} {t-t_0}
\quad ,
\end{equation}
and  Figure~\ref{contraction} 
shows such a contraction 
of the framework which sees the expansion
when the velocity is evaluated 
as a function of time from 
equation~(\ref{velocityrelativistic}).

\section {Conclusions}

The introduction of a relativistic  framework
into the equation of a SNR under the hypothesis 
of the thin layer approximation avoids  
the paradox of an initial velocity greater
than the velocity of light.
The self-similar solutions for radius and velocity, 
respectively eqns. (\ref{radiusrelativisticasyn}) 
and 
(\ref{velocityrelativisticasyn}),
are found under the approximation 
$R^6 (1-\beta_0^2) \gg 
R_0^6\beta_0^2$   .
The application of these new formulae to SNR1987A 
produces acceptable results.
The conservation of classical and relativistic momentum
adopted here 
does not take 
into account the momentum carried away by photons.


\begin{thebibliography}{10}
\expandafter\ifx\csname url\endcsname\relax
  \def\url#1{\texttt{#1}}\fi
\expandafter\ifx\csname urlprefix\endcsname\relax\def\urlprefix{URL }\fi

\bibitem{Israel1966}
W.~{Israel}, {Singular hypersurfaces and thin shells in general relativity},
  {\it Nuovo Cimento B Serie} {\bf 44} (1966), 1--14.

\bibitem{Israel1967}
W.~{Israel}, {Gravitational Collapse and Causality}, {\it Physical Review} {\bf
  153} (1967), 1388--1393.

\bibitem{Nunez1996}
D.~{N{\'u}{\~n}ez}, H.~P. {de Oliveira}, {Dynamics of massive shells ejected in
  a supernova explosion}, {\it Physics Letters A} {\bf 214} (1996), 227--231.

\bibitem{Kijowski2006}
J.~{Kijowski}, G.~{Magli}, D.~{Malafarina}, {Relativistic dynamics of spherical
  timelike shells}, {\it General Relativity and Gravitation} {\bf 38} (2006),
  1697--1713.

\bibitem{Sedov1946}
L.~I. {Sedov}, {Propagation of strong shock waves}, {\it Journal of Applied
  Mathematics and Mechanics} {\bf 10} (1946), 241--250.

\bibitem{Taylor1950}
G.~{Taylor}, {The Formation of a Blast Wave by a Very Intense Explosion. I.
  Theoretical Discussion}, {\it Royal Society of London Proceedings Series A}
  {\bf 201} (1950), 159--174.

\bibitem{Sedov1959}
L.~I. {Sedov}, {\it {Similarity and Dimensional Methods in Mechanics}},
  Academic Press, New York, 1959.

\bibitem{Dalgarno1987}
A.~{McCray}, In: R.~{Dalgarno}, D.~{Layzer} (Eds.), {\it {Spectroscopy of
  astrophysical plasmas}}, {Cambridge University Press}, 1987.

\bibitem{Dyson1997}
J.~E. {Dyson}, D.~A. {Williams}, {\it {The physics of the interstellar
  medium}}, Institute of Physics Publishing, Bristol, 1997.

\bibitem{Padmanabhan_II_2001}
P.~{Padmanabhan}, {\it {Theoretical astrophysics. Vol. II: Stars and Stellar
  Systems}}, {Cambridge University Press}, {Cambridge, MA}, {2001}.

\bibitem{French1968}
A.~P. {French}, {\it {Special Relativity}}, {CRC}, New~York, 1968.

\bibitem{Gradshteyn2007}
I.~S. {Gradshteyn}, I.~M. {Ryzhik}, A. {Jeffrey}, D. {Zwillinger},
  {\it {Table of Integrals, Series, and Products}}, Academic Press, New
  York, 2007.

\bibitem{Abramowitz1965}
M.~{Abramowitz}, I.~A. {Stegun}, {\it {Handbook of mathematical functions with
  formulas, graphs, and mathematical tables}}, Dover, New York, 1965.

\bibitem{Seggern1992}
D.~{von Seggern}, {\it CRC Standard Curves and Surfaces}, CRC, New York, 1992.

\bibitem{Thompson1997}
W.~J. {Thompson}, {\it Atlas for computing mathematical functions},
  Wiley-Interscience, New York, 1997.

\bibitem{Zhang1996}
S. {Zhang}, J. {Jin}, {\it Computation of special functions},
  Wiley-Interscience, New York, 1996.

\bibitem{press}
W.~H. {Press}, S.~A. {Teukolsky}, W.~T. {Vetterling}, B.~P. {Flannery}, {\it
  {Numerical recipes in FORTRAN. The art of scientific computing}}, Cambridge
  University Press, Cambridge, 1992.

\bibitem{Panagia2005}
N.~{Panagia}, {A Geometric Determination of the Distance to SN 1987A and the
  LMC}, In: J.-M. {Marcaide}, K.~W. {Weiler} (Eds.), IAU Colloq. 192: Cosmic
  Explosions, On the 10th Anniversary of SN1993J, 2005, 585--+.

\bibitem{Mitchell2002}
R.~C. {Mitchell}, E.~{Baron}, D.~{Branch}, P.~H. {Hauschildt}, P.~E. {Nugent},
  P.~{Lundqvist}, S.~{Blinnikov}, C.~S.~J. {Pun}, {Detailed Spectroscopic
  Analysis of SN 1987A: The Distance to the Large Magellanic Cloud Using the
  Spectral-fitting Expanding Atmosphere Method}, {\it The Astrophysical Journal} {\bf 574} (2002),
  293--305.

\bibitem{Racusin2006}
J.~L. {Racusin}, S.~{Park}, D.~N. {Burrows}, {Radial Expansion And The X-ray
  Evolution Of SNR 1987A}, {\it Bulletin of the American Astronomical Society}
  {\bf 38} (2006), 330--+.

\bibitem{Park2007}
S.~{Park}, D.~N. {Burrows}, G.~P. {Garmire}, R.~{McCray}, J.~L. {Racusin},
  S.~A. {Zhekov}, {Chandra Observations of Supernova 1987A}, In: S.~{Immler},
  K.~{Weiler}, R.~{McCray} (Eds.), Supernova 1987A: 20 Years After: Supernovae
  and Gamma-Ray Bursters, Vol. 937 of American Institute of Physics Conference
  Series, 2007, 43--50.

\end{thebibliography}

\end{document}